\documentclass[conference]{IEEEtran}
\IEEEoverridecommandlockouts
\usepackage{cite}

\usepackage{textcomp}
\usepackage{xcolor}
\def\BibTeX{{\rm B\kern-.05em{\sc i\kern-.025em b}\kern-.08em
    T\kern-.1667em\lower.7ex\hbox{E}\kern-.125emX}}

\usepackage{amsmath,amssymb,amsfonts}
\usepackage{algorithmic, algorithm}

\usepackage{multicol}
\usepackage{subcaption}
\usepackage{times}  % DO NOT CHANGE THIS
\usepackage{helvet} % DO NOT CHANGE THIS
\usepackage{courier}  % DO NOT CHANGE THIS
\usepackage[hyphens]{url}  % DO NOT CHANGE THIS
\usepackage{graphicx} % DO NOT CHANGE THIS
\usepackage{caption} % DO NOT CHANGE THIS AND DO NOT ADD ANY OPTIONS TO IT
\usepackage{color}
\usepackage{soul}
\usepackage[super]{nth}
\usepackage{url}
\usepackage{xfrac}
\usepackage[font=small]{caption}
\begin{document}

\title{ADDAI: Anomaly Detection using Distributed AI}

\author{Maede~Zolanvari,~\IEEEmembership{Student Member,~IEEE,}
		Ali~Ghubaish,~\IEEEmembership{Student Member,~IEEE,}
        and~Raj~Jain,~\IEEEmembership{Life~Fellow,~IEEE,}% <-this % stops a space
}

\maketitle
\thispagestyle{plain}
\pagestyle{plain}

\begin{abstract}
When dealing with the Internet of Things (IoT), especially industrial IoT (IIoT), two manifest challenges leap to mind. First is the massive amount of data streaming to and from IoT devices, and second is the fast pace at which these systems must operate. Distributed computing in the form of edge/cloud structure is a popular technique to overcome these two challenges. In this paper, we propose ADDAI (Anomaly Detection using Distributed AI) that can easily span out geographically to cover a large number of IoT sources. Due to its distributed nature, it guarantees critical IIoT requirements such as high speed, robustness against a single point of failure, low communication overhead, privacy, and scalability. Through empirical proof, we show the communication cost is minimized, and the performance improves significantly while maintaining the privacy of raw data at the local layer. ADDAI provides predictions for new random samples with an average success rate of 98.4\% while reducing the communication overhead by half compared with the traditional technique of offloading all the raw sensor data to the cloud.
\end{abstract}


\section{Introduction}
\IEEEPARstart{I}{nternet} of Things (IoT) integrated into the Industrial Control Systems (ICSs) is called Industrial Internet of Things or IIoT. IIoT systems are the foundations of the critical infrastructures of a nation. Industrial processes such as smart manufacturing, oil and gas exploration, water distribution and treatment, etc., rely heavily on these systems. Utilizing IoT technology in ICSs enhances network intelligence in the optimization and automation of industrial operations.

Moreover, the impact of Artificial Intelligence (AI) analysis on the rapid growth of the IoT is undeniable. IoT has become a fundamental part of both personal lives and the industrial environments. The integration of AI and IoT automates connections, interactions, and data exchange among machines and devices for comprehensive functionality and higher efficiency without requiring any human in the loop. This automation needs high-speed computations; however, the heavy computational load of AI-based algorithms leaves a burden on the system's data processing speed. This concern is even more troublesome in systems such as IIoT, where real-time operations are essential.

According to the annual report of Stanford University’s AI Index, the speed of AI is outpacing Moore’s Law. Since 2012, AI's required computing power has increased 300000 times compared to the expected seven times increase under Moore’s law \cite{Stanford2019}. The computation power is expected to double every two years based on Moore’s law, while the amount of computing power required for AI doubles every 3.4 months . This proves the unsustainable computing power of AI and the fact that in several years, reaching a general intelligence will not be through a highly powerful complex AI algorithm, but rather through collective intelligence. Collective intelligence means distributing the computation and making decisions through several learning components concurrently and in parallel. This fact emphasizes why distributing the computations in an AI-based IoT system is essential.

%Requiring all the raw data to be offloaded to one centric computing element for processing would cause a bottleneck in the system. In heavy computational scenarios, the created congestion will affect the real-time operations and slow down the critical functions. 

On the other hand, traditional centralized computing is not capable of carrying out, keeping up, and adapting to the required security operations. At the same time, all the sensor data is offloaded into a large and complex AI. It also suffers from an important security vulnerability, i.e., a single point of failure, which negatively impacts the main goal of IIoT systems, high availability and reliability. Also, it comes with high latency, communication costs, and privacy issues. Therefore, our goal in this project is to reduce the need to send all the field data to a central remote computing processor. This will lead to a promising architectural solution to a fundamental network problem, latency.

Despite the increased interest in using rapid computations in AI analysis, most research works focus on distributing the training process.
% usually through updating model parameters. To the best of our knowledge, there has been no research that practically fragments the decision making process of AI models in a granular fashion that would significantly decrease the computation overhead. Not only the training, but also labeling new data should be distributed in a way that their integrity is not compromised.
Our proposed model fragments the decision making-process of AI models in a granular fashion. It can also make decisions locally and later take advantage of those decisions when it comes to classification tasks in the upper layer hierarchy.

%To evaluate and provide a practical instance for \textcolor{red}{NAME}, we use three datasets: 1) ``WUSTL-IIoT," a dataset generated from our prior work on our industrial IoT (IIoT) testbed \cite{Zolanvari2019}; 2) ``NSL-KDD," a public dataset for network-based IDSs from University of New Brunswick \cite{Tavallaee2009}; and 3) ``UNSW," another IDS dataset from the Cyber Range Lab of the Australian Centre for Cyber Security (ACCS) \cite{Moustafa2015}. We empirically prove that our proposed DAI model successfully predicts the labels of unseen data quickly in a matter of few milliseconds. 

The research contributions of this work are as follows: 
\begin{enumerate}
\item We propose a universal, accurate, and well-performing DAI model called ADDAI that does not compromise the learning performance. At the same time, it provides a fair distribution in the computational loads.
%\item We show that \textcolor{red}{NAME} is very fast at classifying the labels of numerical data, making it one of the first DAI models suitable for real-time numerical applications.
\item Our case study introduces a unique perspective on the assessment of AI-based distributed systems for IIoT and Industry 4.0 \cite{Gilchrist2016}. 
\item The proposed framework can be utilized as a low overhead intrusion detection system (IDS) for IIoT systems.
\item Our IIoT security dataset would be released to support the research community for a more extensive and more diverse data collection in the emerging field of AI and DAI for IIoT.
\end{enumerate}

%As data privacy and integrity are essential in all areas of IoT, cybersecurity is of paramount importance. Utilizing AI to provide a secure platform in this area has been proven as an effective method, especially in mission-critical applications such as IIoT \cite{Costa2019}, \cite{Alcaraz2019}, and \cite{Zolanvari2019}. Our case study on the security of IIoT indicates our belief in the importance of IIoT and Industry 4.0. However, \textcolor{red}{NAME} is universal and can be applied to any other applications with minor or zero modification.

\section{Related Work}
DAI is an immense area that includes the distribution of different parts of AI models, such as the datasets, training, tasks, etc. Distribution among different computing entities can be decided based on their differences in energy consumption, memory restrictions, computing power, and many more factors. Additionally, a centralized solution is not even an option when data is inherently distributed or too big to store on a single device. There exists several comprehensive surveys related to this area such as \cite{Ponomarev2017}, \cite{Queiroz2019}, and \cite{Afrin2021}. %, and \cite{Stone2000}. %\cite{Xing2016}

On the other hand, there are situations in which it is beneficial or even required to isolate some subsets of the data. These concerns usually happen when privacy issues are involved. In \cite{Abadi2016}, a framework of differential privacy is considered, and a deep neural network with a modest total privacy loss is developed. Another technique to training models in a privacy-sensitive context is utilizing distributed ensemble models. This guarantees the complete separation of the training data subsets where privacy needs to be preserved. \cite{Sui2020} proposes a federated learning approach that uses ensemble distillation in a medical relation dataset. The suggested technique also overcomes the communication bottleneck caused by the need to upload a large number of parameters in regular federated learning models. Another challenge in these kind of models is that a method needs to be found that properly balances each model’s input for an unbiased training result.

Another popular application of DAI is in unreliable networks, where we have a lossy network and cannot really get a guarantee that sent information will successfully get to the supposed receiver. In \cite{Yu2019}, such scenarios are studied. To overcome this challenge, parameter servers are introduced to store the parameters of the learning model (e.g., the weights of a neural network). These servers serve the parameters to local units (Workers) in charge of processing the data and computing updates to the parameters. Each server connects to all the local units and serves a partition of the model, and each local unit holds a replica of the whole model. However, every communication link between each server and each local unit has a non-zero probability of being dropped.

DAI is a well-known approach in edge-cloud computing. The concept is to move all or some of the training workloads from the central computing unit (e.g., the cloud) to the edges. This way, we reduce the enormous network communication overhead and provide low-latency solutions. To name some challenges in this approach, we can mention parallel training, model synchronization, and workload balancing to address the imbalance of workload and computational power of different edge nodes. For instance, \cite{Chen2019} studies a case study of utilizing DAI in video surveillance systems. Popular services are also cached on the edge servers, so that the latency can be reduced even more, and the computation can be offloaded easily. However, only the cost of bringing services to the edge is considered, and the cost of transmitting data between the edge and cloud server is assumed negligible.

Not all the elements in the IIoT systems have the same computing capabilities. The distribution of the tasks among the elements of the hierarchy must be fair. Therefore, their different computation and communication capabilities should be assessed. The trade-off between cost and capacity of the resources is another important consideration for optimizing resource sharing. There is an extensive study and survey on resource management of different levels of hierarchy and their constraints in \cite{Tocze2018} and \cite{Mijuskovic2021}.

Another approach in the distributed AI is to divide the tasks into micro-services and spread them out in a distributed fashion. These tasks can be data filtering and cleansing, training, testing, labeling, security computation, etc. Different stages of learning development can also be divided into layers among a few high-capacity processing entities. Afterward, the developed model can be accessed simultaneously by several smaller processors to predict the labels of new instances, which is a low computation task. The same mechanism can be applied to other computation tasks. As another example, breaking the training set or the training process into several parallel local tasks can also provide an early result at the local layer and help deal with massive datasets and develop scalable learning. Some examples of research works in this concept include \cite{Li2020}, \cite{Wu2016}, and \cite{Kim2019}.

%The tasks that our produced distributed AI can handle would span from simple anomaly detection to explaining the reasons behind the AI’s actions. Since not all the processing entities have equal computational power, the distributed learners will be designed to be feasible with available processing resources, such as computation, memory, time, and communication. Depending on the type of computations, the tasks can be divided among the sensor’s platforms, the local processors, and the remote processor. Another benefit would be scalability. The produced model would be able to easily spanned geographically.

%The produced model would be able to process most of the data locally with good performance while reducing communication overload. One possible approach is to achieve this by utilizing shallow or simple models when there are capability restrictions or complex deep models otherwise. We would need an optimal data fusion scheme to aggregate outputs of different models into one coherent result with high accuracy.

\section{Background}
ADDAI is composed of two learning models, autoencoders and AdaBoost. For the sake of completeness, we first provide the basics of how these two models function before diving into the details of our proposed model.

\subsection{Autoencoder}
%Autoencoders can work with voluminous unlabeled data and extract useful features with good accuracy. This is specifically useful in industrial systems since a large percentage of data is unlabeled, and it is very time comsuming and labor intensive to be tagging this large amount of data. Autoencoders can be used to extract useful features from high dimensional sensor data. 

Autoencoder is a type of neural network with a bottleneck layer in its network, forcing it to produce a compressed knowledge representation of the original input (also known as the code). The network consists of two parts, an encoder part that produces the code and a decoder that reconstructs the input from the code. By learning a smaller size code, autoencoders can work with voluminous unlabeled data and extract useful features with good accuracies \cite{Goodfellow2016}. %, such as sensor data in industrial systems. Learning a smaller size code forces the autoencoder to capture the most salient features of the input. 

%Autoencoders provide dimension reduction and the output data is laten variables. They remove the redunduncy in the data and combine all the features together rather than just simply pick the important features from the inputs, which is the most common way of dimension reduction.

\begin{figure}[h]
\centering
  \includegraphics[scale=0.46]{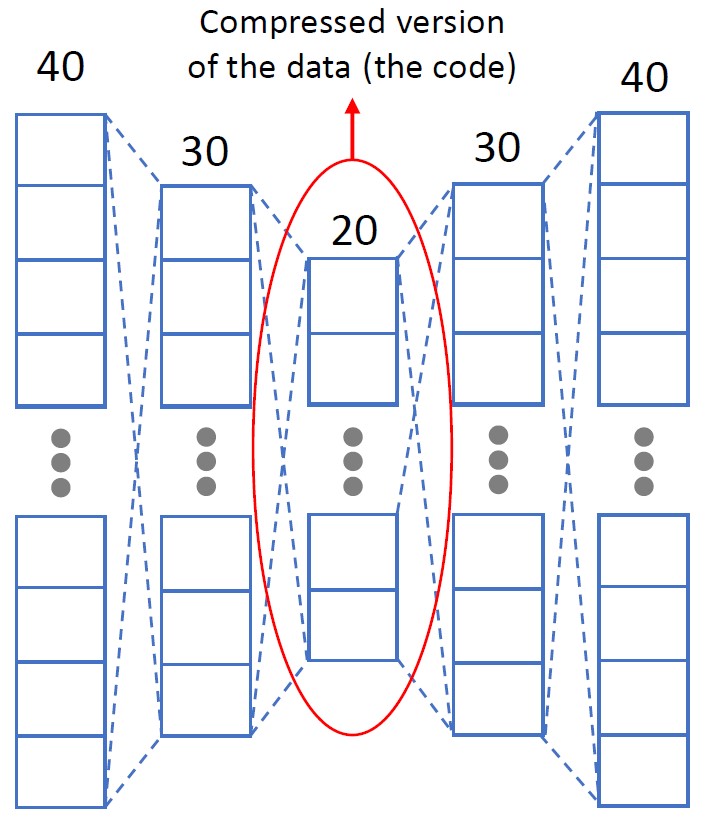}
  \caption{Diagram of an autoencoder. The middle layer, which is called the code, is the compressed version of the input.}
  \label{autoencoder}
\end{figure}

Suppose we have a training set $X\neq \emptyset \in \mathbb{R}^{N\times K}$, where each sample $x_i$ in $X=\{x_1, x_2, ..., x_N\}$ is a $1\times K$ vector (i.e., $K$ features). An example is shown in Fig. \ref{autoencoder}. In this example, we have a dataset in which each sample has $K=40$ features, and the size of the produced code is 20.

\vspace{-.4cm}

Any neural network can be uniquely identified and built by its hyper-parameters. There are four hyper-parameters for an autoencoder which are weights $\boldsymbol{W}$, biases $\boldsymbol{b}$, code size, and the number of hidden layers (when the number of nodes per layer and the loss function are fixed). The weights specify the weighted connections of neurons between every two adjacent layers, while the biases shift the activation function by adding a constant.

Choosing the size of the code and the number of hidden layers both have specific trade-offs. The size of the code should not be too small. First, we might lose more information from the data. Second, it would be tough to capture all the relationships among the input features. This causes underfitting. Meanwhile, the size of the code should not be too large. The network might simply overfit the code layer by learning the noise and small details of input data.

On the other hand, the number of hidden layers should not be too high. The gradient effects of the first layers become too small, and the input becomes almost irrelevant to the code. On the contrary, if the number of hidden layers is too small, it cannot extract useful relations. In neural networks, the first layers usually extract very simple relations like lines and curves, but deeper layers extract more complex features.

Construction error is another important concept in autoencoders. As mentioned before, the decoder part tries to mimic the input layer in the output layer. The more similar these two layers are, the more accurately the autoencoder extracts the code. The most popular way of optimizing the autoencoder's parameters to minimize the construction error is using the mean squared error (MSE), Eq. \ref{mse}

\vspace{-.35cm}
\begin{equation} \label{mse}
\resizebox{.35 \textwidth}{!}
{$L(x_i,{x_i}^\prime) = 1/K \sum_{j=1}^{j=K} {(x_{i,j}- {x_{i,j}}^\prime)}^2$}
\end{equation}

where, $x_i,$ is a sample for the input layer (e.g., the leftmost 40 dimensional layer in Fig. \ref{autoencoder}) and ${x_i}^\prime$ is the output layer (e.g., the rightmost 40 dimensional layer in Fig. \ref{autoencoder}).

\subsection{AdaBoost}
AdaBoost, which is short for Adaptive Boosting, is a popular boosting technique. Boosting is an ensemble method that improves the performance of a number of weak learners and builds a strong learner. Decision trees are the most suited and commonly used weak learners in AdaBoost. In this technique, the weak learners are added sequentially and trained on repeatedly modified versions of the data (i.e., different weights for each instance). Training stops when a pre-defined number of weak learners are trained, or no further improvement can be made. Afterward, the predictions from all the weak learners are combined through a weighted sum or vote to be decided as the final decision, Eq. \ref{adaboost}. %This information infusion is the basis of ensembel learning. Each learner has a stage value that is used to weight its predictions in the final decision on the intances' label. The stage value is simply based on their performance so that the  more accurate models have more contribution to the final prediction.

\vspace{-.35cm}
\begin{equation} \label{adaboost}
\resizebox{.2 \textwidth}{!}
{$L_T (X)=  \sum_{t=1}^T \alpha_t l_t(X)$}
\end{equation}
where $\alpha_t = \frac{1}{2} \ln(\frac{1-\epsilon_t}{\epsilon_t})$ is the weight of $t$'th weak learner, in which $\epsilon_t = (\sum_{y_i \neq l_t(x_i)} w_{i,t})/(\sum_{i=1}^N w_{i,t})$ is its weighted error rate.

In AdaBoost, we initially give each instance in the training dataset equal weights. As the training progresses, at each step, it gives more weights to individual instances that were classified wrong and less to those already correctly classified. The updating rule for the weight of each training sample is based on Eq. \ref{w_update}.

\vspace{-.4cm}
\begin{equation} \label{w_update}
\resizebox{.25 \textwidth}{!}
{$w_{i,t+1} = w_{i,t} \times e^{-y_i l_t(x_i) \alpha_t}$}
\end{equation}

If we assume $y_i \in {-1,1}$, when the prediction of the weak learner is correct (i.e., both $y_i$ and $l_t(x_i)$ are $-1$ or $1$), the coefficient to updating the weight is $e^{- \alpha_t }$. This means we decrease the weight. When they are different, the updating term will be $e^{\alpha_t }$ which means more weight is assigned to a misclassified sample. \footnote{In our dataset, class normal is labeled as $0$, but for the sake of weight updates in AdaBoost, we represent the class normal with $-1$ instead of $0$. This is just a label and does not have any numerical value.}

\section{Proposed Model}
We propose a layer-wise prediction scheme. These layers include local processor units, the edge, and the cloud. The local processor is in direct contact with a group of sensors working together. We assume near zero communication delay between the sensors and the local processor. The edge can be a middle point between the local processors and the cloud for lower latency computations than cloud offloading. For simplicity, here, we consider a two-layer hierarchy, i.e., the local-cloud scenario. %Our proposed method can be easily generalized to a general local-edge-cloud scenario.
In the following, we discuss the formulate the problem and details of the proposed technique.

\subsection{Problem Formalization} %hypothesis 

In our proposed scenario, the tasks are distributed among these components (i.e., local and cloud) for higher efficiency in terms of latency and communication overhead. One important task that is done locally and distributed all over the network is anomaly detection. This way, the anomalies can be detected by the local processing units quickly without the need for sending the data to any other local unit or the cloud. On the other hand, the computation-intensive tasks are done in the cloud. These tasks include training the local models and further investigating the label of the data when we are still uncertain about it despite the local processing.

Two scenarios can be defined whether we can trust the classification of the local model at the local processing unit or not? If yes, local decisions are made, and then only the final results are sent to the cloud. %higher hierarchy (e.g., the edge and/or the cloud).
 If the classifications of the local model cannot be trusted, e.g., due to low accuracy rates, further investigation is required on the sample's label. We design the local units such that at least a compressed version of the data that is much smaller compared to the original raw sensor data that would be sent to the cloud. With this scheme of layer-wise computations, the communication between the local processors and the cloud will be decreased to a significantly lower rate.

%It is important to note that, we suppose that the network data from the local units have approximately the same topology and fromat. Hence, at a given point in time across all the local units, network readings are similar if no anomaly occurs. If this assumption cannot be made and the system is a large-scale network with different protocols and data, the proposed model can easily scale out and group the similar nodes in the network into a cluster. Each cluster would require a tarilored autoencoder and the cloud would maintain all the autoencoders seperately.

\subsection{Architecture}

\begin{figure}[h]
\centering
  \includegraphics[scale=0.5]{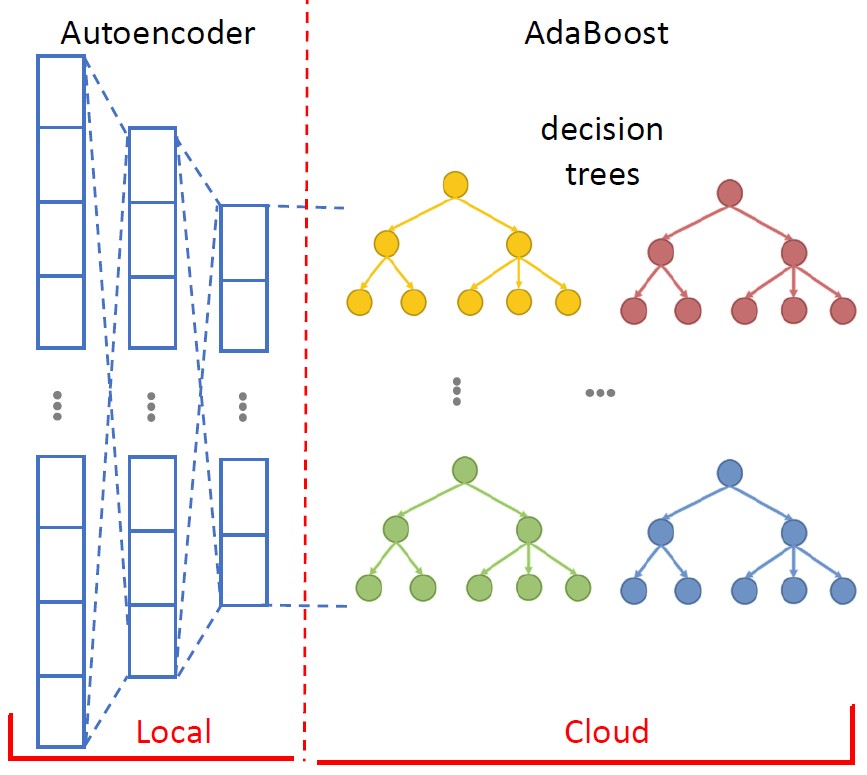}
  \caption{Diagram of the proposed model. The raw data first is fed into a local autoencoder to compress and resize; then, they will be fed into an AdaBoost model in the cloud to decide on their label.}
  \label{diagram}
\end{figure}

We develop an autoencoder as the proper learning framework for local processing units. Fig. \ref{diagram} shows the diagram of the proposed model.
%They are also known as dimension reduction technique, hence, only a more abstract and less redundant information is sent up to the cloud. This helps to reduce the required network communication significantly when it is necessary to send the data to the cloud.
With their produced compressed version of the input, the communication cost significantly drops. This design also guarantees privacy restrictions because latent variables composed from the data features are sent to the cloud instead of the raw data from sensors. 

We train the local autoencoder by normal data only. The achieved model parameter ($\boldsymbol{W}$ and $\boldsymbol{b}$) in the trained autoencoder are used to accurately produce the compressed data (representing code) of the normal samples. Therefore, the compressed data is a model for the normal data. When an attack sample is fed to the model, the reconstruction error is very large, which is how the anomalies are detected. The larger the error is, the more confident we are that the sample is an attack.

Since the training might be computationally heavy for some of the local units, the cloud trains the autoencoder model using the data from all the sensors, $X=\{x_1, x_2, ..., x_N\}$. Here, $N$ is the total number of samples from all the sensors across the network. After the training is done, the cloud sends the updated model parameters $\boldsymbol{W}$ and $\boldsymbol{b}$ back to the local units to build a local version of the autoencoder. Please bear in mind that training is a one-time process.%, and we can assume the features of the data does not change drastically over time, so the trained model is generalized enough and stays good for a long period of time. 

If the predictions of the local processing units are trustworthy, in case an anomaly is detected, they can quickly take action by alarming the control center. However, sometimes the compressed data is sent to the cloud for further investigation. In the cloud, we utilize three different trained AdaBoost models. One of these models is more sensitive on detecting the normal class; another one is sensitive to the attack class, and the third one is a neutral model that pays fair attention to both classes. In other words, these three models are trained in a way that they would have different  sensitivity/recall scores for different classes of the data.  Since AdaBoost works with weights for individual instances, we have assigned class weights at the decision tree levels (through the weak learners). This class weights can also be very helpful when dealing with imbalanced datasets, where the minority class is usually overlooked by the model. The loss function for each of the AdaBoost model can be modeled as in Eq. \ref{loss_cloud}.
\vspace{-.3cm}

\vspace{-.3cm}
\begin{equation} \label{loss_cloud}
\text{loss} = \frac{1}{N} \sum_{i=1}^{N} \! cw_0(y_i \times \log y_i^\prime) +cw_1((1-y_i) \times \log(1-y_i^\prime))
\end{equation}
where $cw_0$ and $cw_1$ are the class weights in the case of binary classification. A grid search is conducted to get the optimal value of these weights.

In our proposed model, despite the fact we might not fully trust the local decisions on the data labels, we want to take advantage of them. We utilize them in a new concept we call \textit{local ranges}. %to identify the boundries for which samples would be fed to which AdaBoost model in the cloud.

\subsection{Local Range}
The local range is a critical component of our proposed architecture. It determines the three ranges of the samples fed to each AdaBoost model in the cloud. It is specific for each local processing unit and depends on how certain we are in the predictions of the samples made by that local unit classifier. A local unit has an accuracy score between $0$ and $1$; we call it $f_i$ for $i$'th local unit. Therefore, the ratio of the number of samples that are misclassified by the autoencoder in the $i$'th local unit is $r_i  = 1 -  f_i$.

On the other hand, when the local unit calculates the reconstruction errors, we can sort them in ascending order. Let us call the sorted array $Err_i$. Local autoencoder would classify any sample with a reconstruction error below a threshold, $\eta_i$, as normal and any sample above that as an attack. With good approximate, we can assume that misclassification happens around the threshold. We want to feed these samples to the regular model as it is partial to both classes. Therefore, we define the range of the reconstruction error of the samples that should be fed to the regular model is defined as

\vspace{-.5cm}
\begin{equation} \label{certainty}
\begin{split}
\{ & \eta_i -  Err_i[idx - \lfloor (r_i \times N_i/2)\rfloor], \\
& \eta_i +  Err_i[idx+\lfloor (r_i \times N_i/2)\rfloor] \}
\end{split}
\end{equation}
where $\eta_i$ is the threshold calculated by the autoencoder for the $i$'th local unit based on its pick performance on the training data specific to that local unit. $idx$ is the sample index whose reconstruction error is the closest to the value of $\eta_i $, and $N_i$ is the number of training samples in the local unit $i$.

%reconstruction error of the sample in the local unit. It is important to note that while $f_i$ can be defined as a number between 0 and 1, the reconstruction error can take any value greater or equal to zero. However, we suggest to clip the error values to a pre-defined maximum value equal to the maximum error from the validation set. This way, we contain the values of $e_j$s.

\begin{figure}[h]
\centering
  \includegraphics[scale=0.36]{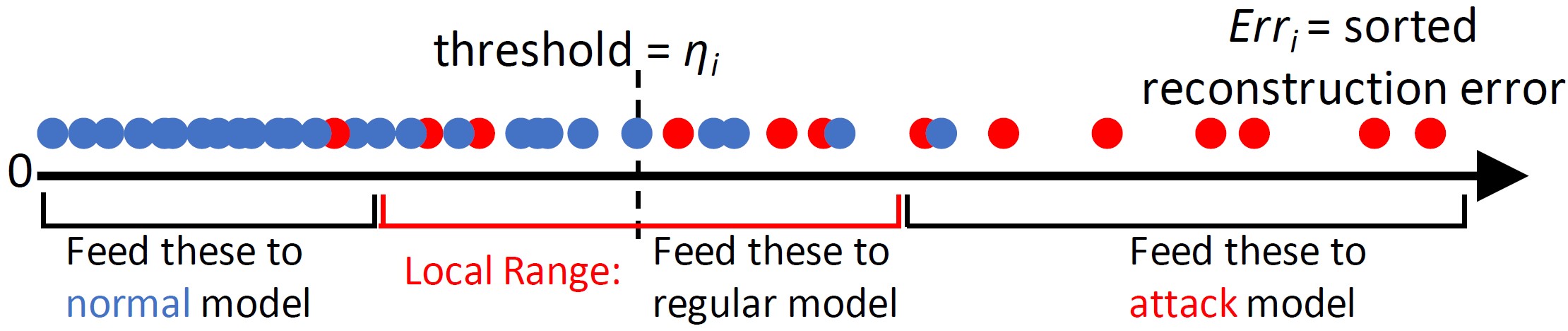}
  \caption{Local decisions based on the reconstruction errors from the autoencoder.}
  \label{local_decision}
\end{figure}

As shown in Fig. \ref{local_decision}, we make the decision on which sample to be fed to which model based on the local range and the value of the reconstruction error of the sample. The samples with reconstruction errors within the local range to be fed to the regular model, and below or above that to the normal and attack models, respectively. This figure shows an example of reconstruction error for a set of test samples. The more uncertain we are about the local decisions, the wider the boundary for the regular model is.

\vspace{-.5cm}

%Here is a  step by step summay of of my model:
%\begin{enumerate}
%\item all the local units send their training data to the cloud.
%\item cloud trains a model on all the collected training data, sends the parameters to the locals.
%\item locals use the model and their own training subset to pick the best error threshold for their data.
%\item test set (which is unseen to any processing unit) comes in, they use the local model to decide on the label and compress the data to be sent to the cloud.
%\end{enumerate}

\vspace{-.5cm}
\subsection{Communication Cost}

%Note that the intermediate output can be designed to be much smaller than the sensor input (e.g., a raw image from a video camera), and therefore drastically reduce the network communication required between the end device and the cloud.
The communication cost for offloading the data to the cloud, $k$, can be formulated as follows:
\begin{equation} \label{comm}
\resizebox{.4 \textwidth}{!}
{$k = 1 \text{ (for $\mathcal{C}$)} + 4 \text{ (for reconstruction error)} + 4 \times h$}
\end{equation}

In this equation, $\mathcal{C}$ is the predicted class by the local unit, and one byte is assigned for it. 4 bytes are dedicated for the reconstruction error. $h$ is the length of the code layer of the autoencoder. 4 bytes are used to represent a signed single-precision floating-point number.

\section{Case Study: IIoT Security }
%Sophisticated high computational AI models are vastly utilized in industrial settings. Intrusion detection systems (IDS) for network security is one of the common applications in which AI is a popular approach . Utilizing \textcolor{red}{NAME} would distribute and speed up the detection process in these systems. 
We have built a lab-scale industrial Internet of Things (IIoT) system to collect realistic and up-to-date datasets for network-based intrusion detection systems (IDS). We have implemented a supervisory control and data acquisition (SCADA) system widely used by industries to supervise the level and turbidity of liquid storage tanks. This IIoT system is employed in industrial reservoirs and water distribution systems as a part of the water treatment and distribution. More information regarding our testbed can be found in our previous papers \cite{Zolanvari2019} and \cite{Teixeira2018}.

\vspace{-.1cm}
%An essential step in training the learning models is selecting and extracting features from the traffic.
\subsection{Utilized Datasets}
The dataset is the one collected from our testbed, which we refer to as ``WUSTL-IIoT."  To collect the proper dataset, we (as a white-hat attacker) attacked our testbed with different cyber-attacks \cite{Zolanvari2019}. %Here, we label  all the attack data as class 1 for binary classification.
Specifics of our dataset are in Table \ref{three-DS}.

\begin{table}[h]
\centering\fontsize{8}{9}\selectfont
\caption{Specifics of the tested dataset}
\begin{tabular}{|c|c|}
\hline
\textbf{Dataset} & WUSTL-IIoT  \\
\hline
\textbf{\# of observations} & 1,194,464 \\
\hline
\textbf{\# of features} & 41 \\
\hline
\textbf{\# of attacks} & 87,016 \\
\hline
\textbf{\# of normals} &1,107,448  \\
\hline
\end{tabular}
\label{three-DS}
\end{table}

\vspace{-.3cm}
\subsection{The Learning Models}
The inputs to the learning models are the flow instances, as mentioned in the previous subsection. The output of the models can be a multi-class or binary classification. However, for simplicity, we have built ADDAI by treating the outputs of the model as binary classes with normal as 0 and attack as 1. 

The autoencoder used in the local processing is programmed using the neural network module of PyTorch \cite{pytorch}, containing the encode, decode, and forward methods. Depending on the code size, we use a different number of hidden layers. For instance, for a code size of 25, 7 layers are used in the model: input data layer, two encoding layers, code layer, two decoding layers, and the output layer, as shown in Table \ref{autoencoder-hp}.
% after the first layer (i.e., the input layer), it has two hidden layers with 35 and 30 neurons. A hundred epochs with \textit{adam} optimizer and a learning rate of $0.01$ and the dropout regularization technique with $0.05$ rate to avoid overfitting are utilized. For the activation function, we have used hyperbolic tangent, or tanh, function, which is defined as $\tanh(x) = (e^{x}-e^{-x}/(e^{x}+e^{-x})$. 
AdaBoost is used in the cloud with decision tree classifiers as the weak learner. The number of estimators in AdaBoost was set at 100. The scikit-learn library \cite{Pedregosa2011} was used to implement the AdaBoost models.

\begin{table}[h]
\centering\fontsize{8}{9}\selectfont
\caption{Local autoencoder model hyperparameters}
\begin{tabular}{|ll|}
\hline
\textbf{Parameter} & \textbf{Typical value(s)}  \\
\hline
\# of layers & 7 \\
\hline
\# of neurons per layer  & 40, 35, 30, 25, 30, 35, 40  \\
\hline
\# of epochs & 100 \\
\hline
Optimizer & \textit{adam} \\
\hline
Dropout rate & 0.05 \\
\hline
Learning rate & 0.01 \\
\hline
Activation function & hyperbolic tanh function \\
\hline
\end{tabular}
\label{autoencoder-hp}
\end{table}

\vspace{-.2cm}
\subsection{Performance Metrics} \label{metrics}
In this paper, we evaluate the model's performance using three metrics, including Accuracy, MCC, and undetected rate (UR) shown in Eq. \ref{accu},  Eq. \ref{mcc}, Eq. \ref{ur}, respectively.

\vspace{-.4cm}
\begin{equation} \label{mcc}
\resizebox{.4 \textwidth}{!}
{$\text{MCC} = \frac{TP \times TN - FP \times FN}{\sqrt{(TP+FP)(TP+FN)(TN+FP)(TN+FN)}}$}
\end{equation}

\vspace{-.3cm}
\begin{equation} \label{accu}
\resizebox{.25 \textwidth}{!}
{$\text{Accuracy} = \frac{TP+TN}{TP+TN+FP+FN}$}
\end{equation}

\vspace{-.3cm}
\begin{equation} \label{ur}
\resizebox{.13 \textwidth}{!}
{$\text{UR} = \frac{FN}{FN+TP}$}
\end{equation}
%
%\begin{equation} \label{f1}
%\resizebox{.18 \textwidth}{!}
%{$\text{F1} = \frac{TP}{TP+\frac{1}{2}(FP+FN)}$}
%\end{equation}

\noindent where $TN$ is the number of normal data labeled as normal, $TP$ is the number of attack data classified as an attack, $FP$ is the number of normal labeled as an attack, and $FN$ is the number of attacks labeled as normal by the model.

\vspace{-.1cm}
\subsection{Results} \label{results}
\begin{figure}[h]
\centering
  \includegraphics[scale=0.41]{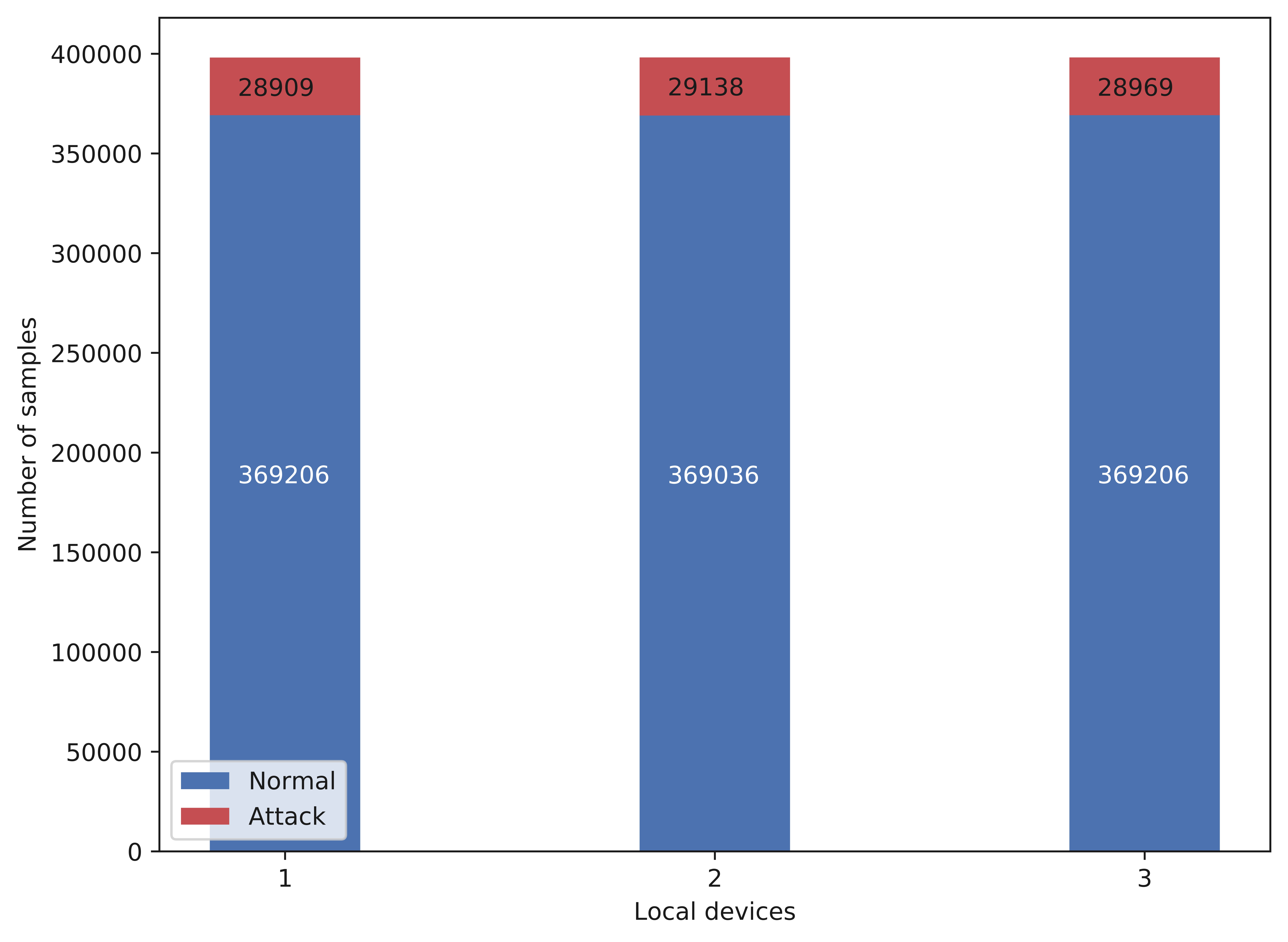}
  \caption{Number of samples from each class of data (Attack or Normal) at each local device.}
  \label{num_samples}
\end{figure}

\vspace{-.1cm}
In this experiment setup, we set the number of local units to be $3$. We have randomly divided the WUSTL-IIoT dataset into three equal sized subsets. Fig. \ref{num_samples} shows the exact number of samples of each class per local device.

\vspace{-.3cm}
\vspace{-.3cm}

Here, we test the performance of ADDAI assuming the worst-case scenario, where we cannot trust any of the local decisions on the samples' label, and they should be sent to the cloud for further investigation. However, we take advantage of the local analysis as mentioned before.

\begin{figure}[h]
\centering
  \includegraphics[scale=0.45]{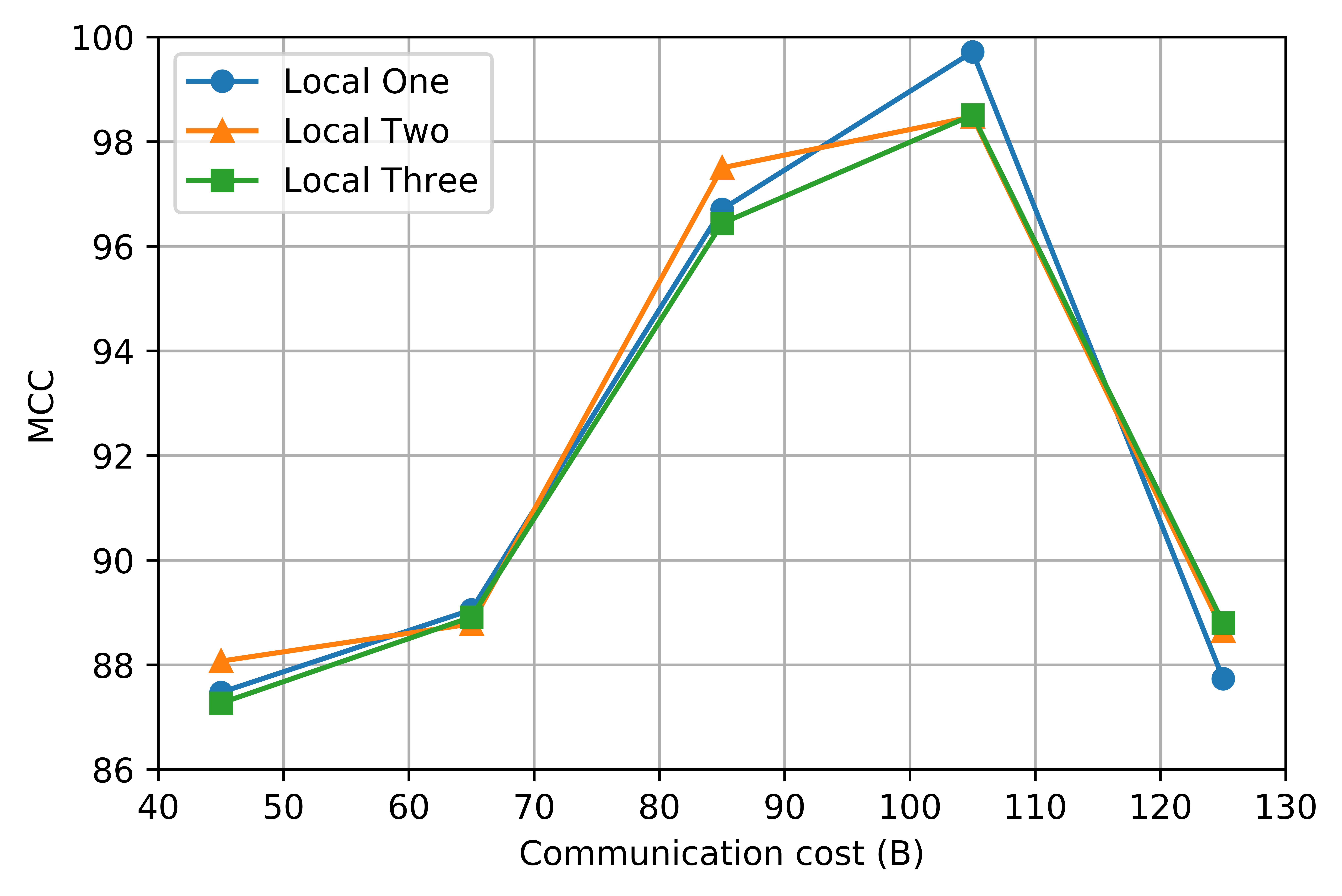}
  \caption{Communication cost vs. MCC scores for different code lengths.}
  \label{comm_cost}
\end{figure}

Our analysis starts by choosing the right code size based on the performance and the communication cost. For different values of $h$, $10, 15, 20, 25,$ and $30$, the associated communication cost (Eq. \ref{comm}) and the performance of the local devices are compared. Fig. \ref{comm_cost} represents the change in the MCC results at the cloud for each of these values. The code size of $25$ with an average of $105$B communication cost produced the best result. Therefore, in the rest of the evaluations, we use the code size of $25$ unless stated otherwise. For this setting, the stored autoencoder models on the local processing units require only 16 kB of memory.

%So, the number of dimensions of the data start at $41$ and would be shrunk to $25$ before being sent to the cloud. As an example, for local device one, this will cost us a total of 105 B*78835 = 8277675 B about 8.3 MB if data to be transferred. While, if I choose to go with a code size of 20, that will cost 85 B per sample that requires 6700975 B about 6.7 MB for a slightly degraded performance, and here it is where the trade-off shows up. Let’s assume a 419 kbit/s which is about 52.37 kB/s data rate (as this is the data rate of our PLC in our test bed) going to local as the input. Each sample has 41 features, as aforementioned calculation, we dedicate 4 B for each feature, that makes each sample to be around 164 B. In my DAI scenario, we would require sending only 85 B, which is half. The number of samples per second is about equal to 319 samples, which the proposed DAI can easily handle since it requires only half of the bandwidth, about 26.18 kB/s.

\vspace{-.2cm}

\begin{figure}[h]
\centering
  \includegraphics[scale=0.36]{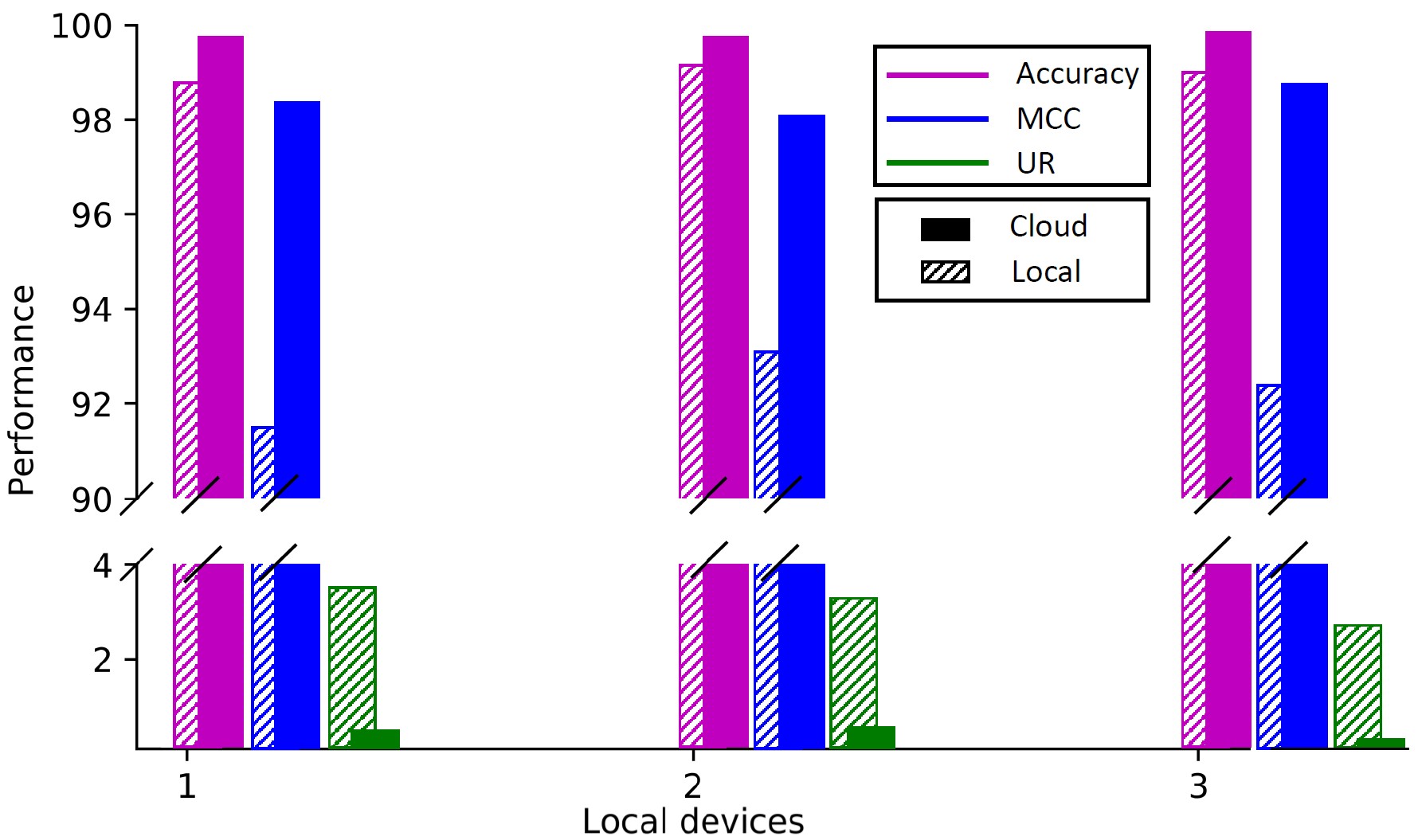}
  \caption{Different performance metrics comparison of the three local devices vs. their corresponding cloud results. }
  \label{performance_metric}
\end{figure}

To get a concrete comparison of the performance improvement through our model, the performance metrics mentioned in Section \ref{metrics} calculated from the values in the previous tables are summarized in Fig. \ref{performance_metric}. Please note that, for accuracy and MCC metrics, the higher the value, the better, while for UR, the lower the value, the better. %Table \ref{AI_summ}.
Each local unit splits its dataset into training and test sets with an 80:20 ratio. Since the autoencoder in the local units predicts the labels (regardless we can trust them or not), we have utilized them as the local performance on the data. The cloud results are derived from feeding different samples to different AdaBoost models (regular, normal, and attack) based on their calculated local ranges. Table \ref{local_ranges} shows the utilized ranges for each local device. These numbers are calculated using Eq. \ref{certainty}.

%\begin{table}[H]  
%  \centering\fontsize{9}{11}\selectfont
%  \caption{Summary of the local and cloud performances in all three datasets}
%  \begin{tabular}{|p{3cm}|p{1.2cm}|p{1cm}|p{.8cm}|}
%    \cline{2-4}
%    \multicolumn{1}{c|}{} & \textbf{Accuracy}  & \textbf{MCC} & \textbf{UR}\\ \hline
%    \mbox{One, local} & 98.76\%  & 91.36\% & 3.44\% \\ \hline
%    One, cloud  & 99.78\%  &  98.39\% & 0.34\% \\ \hline
%    Two, local  & 99.01\%  & 93.03\%  & 3.29\%\\ \hline
%	Two, cloud  & 99.73\%  &  98.07\% & 0.51\%\\ \hline
%	Three, local  & 98.91\%  & 92.38\%   & 2.69\% \\ \hline
%	Three, cloud  & 99.82\%  &  98.74\%  & 0.14\%\\ \hline
%  \end{tabular}
%   \label{AI_summ}
%  \end{table}
\vspace{-.3cm}

\vspace{-.4cm}
\begin{table}[H]
\centering\fontsize{8}{9}\selectfont
\caption{Specifics of the tested dataset}
\begin{tabular}{|c|c|c|}
\hline
\textbf{Local device} & $\boldsymbol{\eta_i}$  & \textbf{Normal range} \\
\hline
One &  0.43 & [0.29, 1.38] \\
\hline
Two&  0.58 & [0.17, 1.43] \\
\hline
Three & 0.42 &  [0.28, 1.35] \\
\hline
\end{tabular}
\label{local_ranges}
\end{table}

\begin{figure}[h]
\centering
  \includegraphics[scale=0.46]{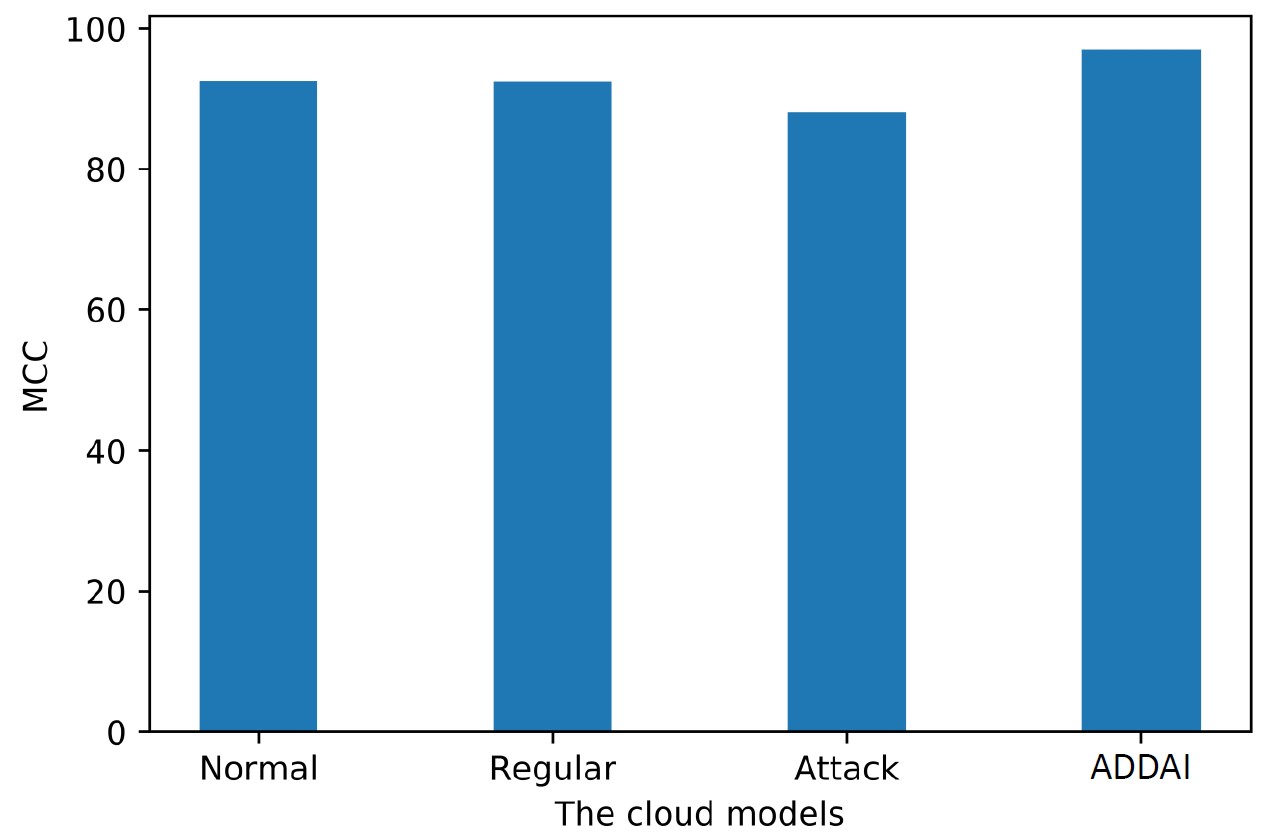}
  \caption{The MCC results, if only one of the cloud models were utilized.}
  \label{comp_AdaBoost}
\end{figure}

\vspace{-.3cm}
We also tested what would happen if we feed all the samples to only one of the AdaBoost models in the cloud. For simplicity, we combined the data from all the local devices. The results are shown in Fig. \ref{comp_AdaBoost}.%Table \ref{feedToOne}.

%\begin{table}[H]  
%  \caption{The MCC results, if only one of the cloud models was utilized.}
%  \begin{subtable}{.5\linewidth}
%  \centering\fontsize{9}{11}\selectfont
%  \begin{tabular}{|c|c|c|}
%    \cline{2-3}
%    \multicolumn{1}{c|}{} & Normal & Attack \\ \hline
%    Normal& 220213 &  1243   \\ \hline
%    Attack & 1198  & 16239     \\ \hline
%	\multicolumn{3}{|c|}{MCC =  92.45\%}  \\ \hline
%  \end{tabular}
%  \caption{Regular model}
%   \end{subtable}%
%    \begin{subtable}{.5\linewidth}
%      \centering\fontsize{9}{11}\selectfont
%    \begin{tabular}{|c|c|c|}
%    \cline{2-3}
%    \multicolumn{1}{c|}{} & Normal & Attack \\ \hline
%   Normal& 220293  & 1163   \\ \hline
%    Attack & 1250  & 16187    \\ \hline
%\multicolumn{3}{|c|}{MCC =  92.51\%}  \\ \hline
%  \end{tabular}
%   \caption{Normal model}
%   \end{subtable}
%\begin{subtable}{.5\linewidth}
%  \centering\fontsize{9}{11}\selectfont
%  \begin{tabular}{|c|c|c|}
%    \cline{2-3}
%    \multicolumn{1}{c|}{} & Normal & Attack \\ \hline
%    Normal& 216932 &  4524   \\ \hline
%    Attack & 26  & 17411     \\ \hline
%\multicolumn{3}{|c|}{MCC =  88.10\%}  \\ \hline
%  \end{tabular}
%  \caption{Attack model}
%   \end{subtable}%
%    \begin{subtable}{.5\linewidth}
%      \centering\fontsize{9}{11}\selectfont
%    \begin{tabular}{|c|c|c|}
%    \cline{2-3}
%    \multicolumn{1}{c|}{} & Normal & Attack \\ \hline
%   Normal& 220558  & 898   \\ \hline
%    Attack & 112    & 17325    \\ \hline
%\multicolumn{3}{|c|}{MCC =  96.97\%}  \\ \hline
%  \end{tabular}
%   \caption{Our proposed model}
%   \end{subtable}
%   \label{feedToOne}
%  \end{table}

\vspace{-.3cm}

\vspace{-.3cm}
As seen from these results, our evaluations show taking advantage of our prior knowledge of the label of the data helped our proposed model achieve better performance.

\section{Conclusion}
DAI has proven beneficial in applications such as IIoT. In this paper, we have proposed a low overhead DAI called ADDAI. In our proposed framework, anomaly detection is done close to the sensor level, while more investigation on the data can be done in the cloud. To conserve the communication resources, we send a compressed version of the data to the cloud. Through our proposed models, we also preserve the privacy requirements by sending a latent variant of the data to the cloud. We show empirical proof of performance improvement and decreased communication cost of our proposed technique.

\vspace{-.1cm}
\section*{Acknowledgement}
This work has been supported under the NSF grant CNS-1718929. The statements made herein are solely the responsibility of the authors.

\vspace{-.3cm}
\bibliographystyle{IEEEtran}
\bibliography{biblo_MZ}

\end{document}